# Two-Dimensional MoS$_2$ Negative Capacitor Transistors for Enhanced (Super Nernstian) Signal-to-Noise Performance of Next-generation Nano Biosensors

Nicolò Zagni[1,2,†], Paolo Pavan[2], Muhammad A. Alam[1,*]

[1] *School of Electrical and Computer Engineering, Purdue University, West Lafayette, Indiana 47907, USA*

[2]*Department of Engineering "Enzo Ferrari", University of Modena and Reggio Emilia, Modena, 41125, Italy*

**ABSTRACT**

The successful detection of biomolecules by a Field Effect Transistor-based biosensor (BioFET) is dictated by the sensor's intrinsic Signal-to-Noise Ratio (*SNR*). The detection limit of a traditional BioFET is fundamentally limited by biomolecule diffusion, charge screening, linear charge to surface-potential transduction, and Flicker noise. In this paper, we demonstrate that the recently introduced transistor technology called Negative Capacitor Field effect transistor (NCFET) offers nonlinear charge transduction and suppression of Flicker noise to dramatically improve the *SNR* over classical Boltzmann sensors. We quantify the *SNR* improvement by interpreting the experimental results associated with the signal and noise characteristics of 2D MoS$_2$-based transistors. The combined sensitivity enhancement and noise rejection guarantee a high *SNR* of the NC-BioFET, making this device a promising candidate for realizing advanced integrated nanobiosensors.

Many research groups worldwide are focused on developing highly sensitive, fast responding, and selective transistor-based biosensors. These sensors, known as BioFETs, allow label-free detection of biomolecules with the Limit of Detection (LOD) in the nano- and pico- molar concentration ranges[1], potentially enabling many applications in personalized medicine[2], early detection of diseases[3], genome sequencing[4,5], etc. A BioFET relies on the modulation of surface potential due to the charged biomolecules adsorbed on the gate electrode. Thus, the fundamental sensitivity limit of BioFETs depends on the diffusion of biomolecules[6], surface charge screening, and linear "charge-to-surface potential" relationship[7]. These limitations apply to all FET-based biosensors, e.g. Dual-Gate FET[8]; Silicon Nanowire (Si-NW)[1,7,9]; coupled Nanoplate-Nanowire (NW-NP)[10], 2D-semiconductors FET[11–13]. In contrast, electro-mechanically actuated biosensors obviate the screening effect and enhance the sensitivity and signal-to-noise ratio[14] through their inherent nonlinear "mass-vs-deflection" response[15–17]. Unfortunately, the sensitivity improvement is counterbalanced by the difficulty of integrating these sensors into low-voltage CMOS integrated circuits[18,19]. Moreover, sensors based on dynamic mechanical response must be driven by complex oscillator circuits, further exacerbating the integration challenge.

In this paper, we show that a recently proposed, CMOS-compatible nonlinear device called negative-capacitor field-effect transistor (NCFET) can enhance biosensors sensitivity beyond the Nernst limit and thus would be an attractive alternative to the nonlinear electro-mechanical counterparts. The NCFET was originally proposed to reduce the sub-threshold slope (and the power dissipation) of the Metal-oxide field-effect transistors (MOSFETs)[20,21]; however, the inherent non-linearity of NCFETs – arising from the nonlinear negative derivative of polarization vs. field characteristics of a ferroelectric capacitor – can also be exploited to enhance the sensitivity of biosensors. Most importantly, NCFETs can improve the Signal-to-Noise ratio (*SNR*) compared to traditional MOSFETs by reducing the low frequency flicker noise related to carrier number fluctuations[22]. Finally, since NCFETs are fabricated by a simple process-modification that converts standard $HfO_2$ into a ferroelectric film[23,24], they should be easily integrated into the standard CMOS process flow for realizing nanoscale biosensors.

In this letter, we propose the concept of the NC-BioFET combining the NCFET with the conventional BioFET to obtain improved sensitivity and enhanced *SNR*. To this end, we develop a compact model of the NC-BioFET to perform simulations from a circuit perspective. The results of the analysis show that despite the fundamental limits of charged-based BioFETs[8,10], the NC-BioFET can improve the limits of label-free detection of biomolecules. Although we focus on nanobiosensing, the principle of negative capacitance based detection is general and can improve, for example, the *SNR* of potentiometric transistor-based gas sensor[25].

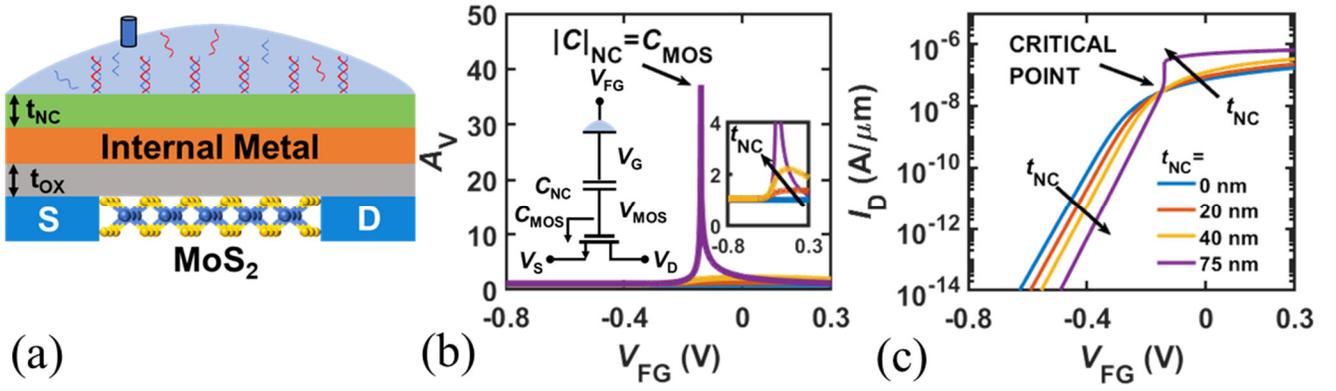

**Figure 1.** (a) Cross-section schematic representation of the NC-BioFET under study. Note that the electrolytic solution lies on the ferroelectric layer. (b) Voltage gain $A_V$ vs. $V_{FG}$ shows a peak when the matching condition is satisfied, i.e., $|C_{NC}| = C_{MOS}$ (symbols are explained in the circuit schematic). The inset shows the increase in $A_V$ with increasing $t_{NC}$. (c) $I_D - V_{FG}$ characteristics for different $t_{NC}$ showing the abrupt transition at the critical point when the $t_{NC}$ is thick enough.

The physical model developed in this work is divided in three fundamental phenomena: *i*) the screening of the electrolyte solution (representing the 'sensing' part of the device); *ii*) the nonlinear response of the negative capacitor and; *iii*) electrical response of the transistor. This modeling approach integrates the negative capacitor with the conventional MOSFET and with the electrolyte model, thereby simplifying the design of NC-BioFETs. Model details are summarized in the supporting information. We will illustrate the principle of operation of the NC-BioFET by using a Molybdenum Disulfide (MoS$_2$) field-effect transistor. The choice of this particular class of transistors is motivated by the investigation of possible biosensing applications for MoS$_2$ FETs, because of their potential for integration in ultra-scaled devices and excellent electrostatic control[12,22].

Fig. 1(a) shows the BioFET configuration involving a MoS$_2$ FET, with the negative capacitor layer included in the gate stack of the MOSFET. Typically, the layer of the MoS$_2$ FET directly exposed to the fluidic environment is the gate oxide layer[12,13], or the channel itself[11]. Similarly to the former case, in this work we consider the ferroelectric as the layer exposed to the fluidic environment. The metal gate is replaced by the electrolyte and by the reference electrode as in common BioFET configurations. The electrolyte is modeled starting from the derivation in ref.[26] for a pH sensor. The effect of charged biomolecules, $Q_{BIO}$, is included in the charge neutrality equation as following:

$$Q_{dl} + Q_{surface} + Q_{BIO} + Q_{MOS} = 0 \qquad (1)$$

where $Q_{dl}$ is the double-layer charge, $Q_{surface}$ is the surface charge due to screening effects and $Q_{MOS}$ is the MOSFET charge. In a first-order approximation $Q_{MOS}$ can be neglected as $Q_{surface}$ almost entirely neutralizes $Q_{dl}$ and $Q_{BIO}$[26]. This phenomenon is at the origin of the so-called screening limited response of FET-based biosensors[7]. Site-binding model parameters[27] are taken by considering HfO$_2$ as the interfacial insulator with the electrolyte. The model takes also into account the screening due to salt ions in the solution and the additional potential drop due to the finite size of charges building up at the interface with the oxide[26,28] (i.e., the so-called Stern layer).

To model the non-linear negative capacitor, we use the well-known phenomenological Landau-Khalatnikov (LK) equation[20,29]:

$$V_{NC}/t_{NC} = 2\alpha Q_{MOS} + 4\beta Q_{MOS}^3 + 6\gamma Q_{MOS}^5 + \rho \frac{dQ_{MOS}}{dt} \qquad (2)$$

where $t_{NC}$ is the negative capacitor thickness and, $\alpha, \beta, \gamma, \rho$ are material and process specific parameters[30,31]. The ferroelectric material under consideration is Zr- doped HfO$_2$ (i.e., HZO). The parameter values used for the simulations can be found in the supporting information. Note that our analysis on sensitivity is focused on DC measurements, therefore the last term in eq. (2) need not be considered. However, this term is important in the AC/noise analysis to account for the ferroelectric contribution to the overall power spectral density, as explained later in the discussion. The NCFET is modeled by following the approach found in ref.[30], considering the negative capacitor and the underlying MOSFET as separate entities solved self-consistently. This approach can be adopted when the internal metal layer is considered in the gate stack, as in this case [see Fig. 1(a)].

The negative capacitor provides a voltage amplification between the internal MOSFET potential $V_{MOS}$ and the applied gate voltage $V_G$. This voltage gain is defined as:

$$A_V \equiv \frac{dV_{MOS}}{dV_G} = \frac{|C_{NC}|}{|C_{NC}| - C_{MOS}} \tag{3}$$

with the symbols of the schematic circuit defined in Fig. 1(b). The increased sensitivity reflects the cancellation of the ferroelectric negative capacitance by the equivalent MOSFET positive capacitance, i.e., $|C_{NC}| = C_{MOS}$[29]. In this condition, the voltage gain $A_V$ exhibits a strong peak, see Fig. 1(b). The peak occurs only if the negative capacitor layer ($t_{NC}$) is sufficiently thick to compensate the traditional dielectric, as it can be seen in the $I_D - V_{FG}$ characteristics, Fig. 1(c). In order to probe the instability region, measurements should be taken by sweeping the Front Gate voltage, $V_{FG}$, (i.e., the potential applied to the reference electrode) for different biomolecules concentration. With increasing $t_{NC}$, the drain current switches abruptly from the off- to the on- state as a consequence of improved capacitance matching. For the parameters considered in this work, a negative capacitance layer of at least 75 nm is necessary to cause an appreciable instability. Note that the device simulated with $t_{NC} = 0$ nm is equivalent to a conventional BioFET with no negative capacitor and with the sensing layer on the 'inner' oxide layer (i.e., without the negative capacitance and internal metal layers).

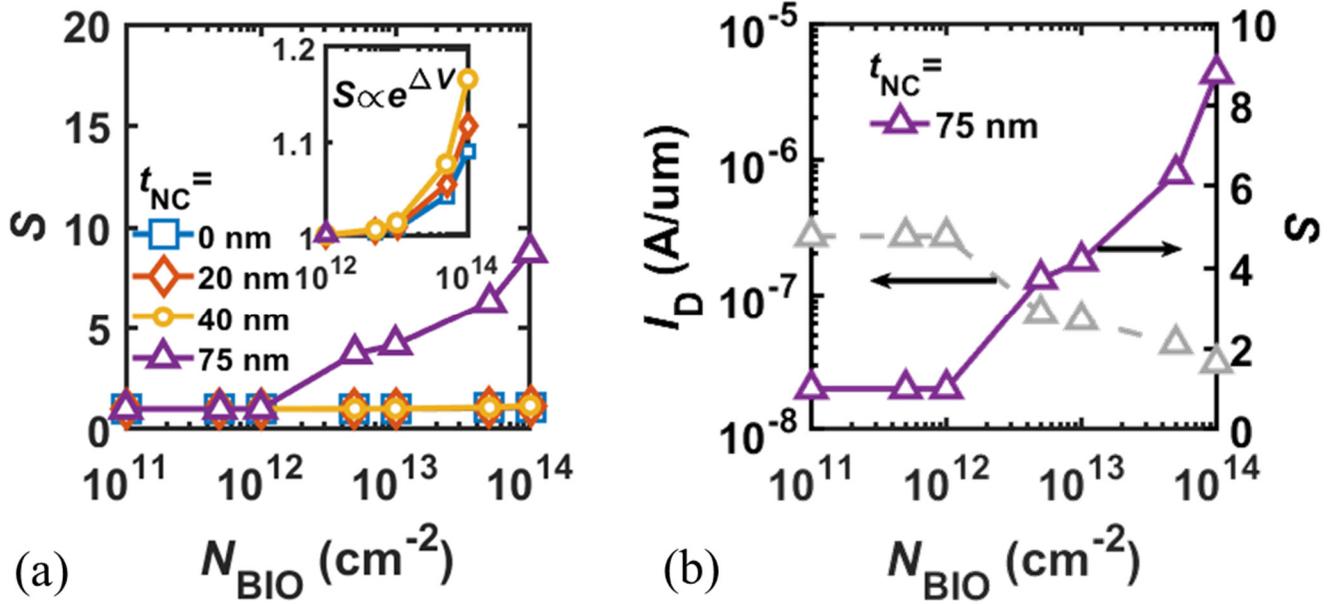

**Figure 2.** (a) Sensitivity vs. biomolecule concentration $N_{BIO}$ with different $t_{NC}$. The sensitivity suddenly increases due to the current transition from inversion to sub-threshold region for the $t_{NC} = 75$ nm case. This transition is shown clearly in (b). The inset in (a) shows a magnification of the plot, indicating that if $t_{NC}$ is not thick enough the sensitivity follows an exponential dependence on the potential shift caused by biomolecules, as one expects in the sub-threshold regime.

The instability at the critical point leads to the increase in sensitivity, as shown in Fig. 2(a). Sensitivity is defined as the ratio between the current before and after the capture of biomolecules, i.e.: $S = I_{before}/I_{after}$. Here, we consider negatively charged biomolecules (e.g., DNA) that increase the threshold voltage and shift the I-V characteristics to the right. The shift in the drain-current due to increasing concentration of captured biomolecules ($N_{BIO}$) triggers the abrupt transition from inversion to sub-threshold operation, thereby producing a significant change in drain current, see Fig. 2(b). As mentioned previously, measurements should be taken by sweeping $V_{FG}$, thus spanning the voltage range where the instability takes place ensuring that enhanced sensitivity is obtained. As expected, the response of the NC-BioFET is indistinguishable from a conventional BioFET as long as the instability is absent (i.e., for low $N_{BIO}$), see the inset in Fig. 2(a). This can be understood by the fact that the screening is still the limiting factor for the response to capture of biomolecules[10].

We now discuss the reduction in noise of the NC-BioFETs. To evaluate the overall noise of the system, we consider the noise sources separately and obtain the output power spectral density by summing up the single contributions (assuming statistical independence between the individual processes). To avoid confusion with the symbol for sensitivity $S$, we refer to the noise power spectral density with $PSD$. The electrolytic solution contributes with thermal noise due its finite conductivity[26]. The negative capacitor contributes to the overall $PSD$ with thermal noise associated with dissipative process due to the damped ferroelectric switching[32], and can be quantified from the last term in Eq. (2). The transistor contributes both thermal and flicker ($1/f$) noise[33]. The thermal noise of the MOSFET is caused by the carriers flowing in a resistive-like channel, whereas Flicker noise arises due to the fluctuation of the number and mobility of carriers in the channel due to the occurring of trapping/detrapping events at the interface with the gate oxide. The general model for the flicker noise in MOSFETs[34] must be modified to obtain an expression that takes into account the negative capacitance effect, i.e.:

$$PSD_{I_D,1/f} = \frac{q^2 \lambda k_B T N_t}{WLC_{eq}^2}\left(1+\alpha\mu_{eff}C_{eq}^2\frac{I_D}{g_m}\right)^2 g_m^2 \qquad (4)$$

where $q$ is the elementary charge, $k_B$ is the Boltzmann constant, $T$ is the temperature, $\lambda$ is the tunneling length, $N_t$ is the oxide traps density, $C_{eq}$ ($C_{eq}^{-1} = C_{NC}^{-1} + C_{ox}^{-1}$) is the equivalent oxide capacitance, $g_m$ is the transconductance, $\mu_{eff}$ is the carriers' effective mobility, and $\alpha$ is the Coulomb scattering coefficient.

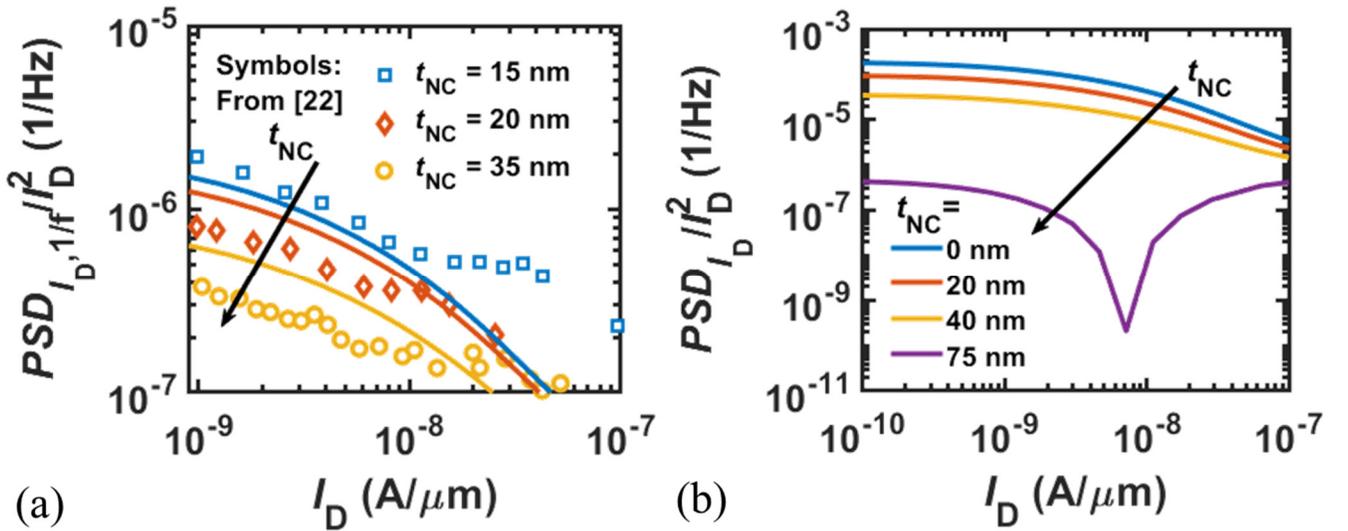

**Figure 3.** (a) Comparison of model from Eq. (4) with experimental data[22] by using $C_{eq} = C_{ox} \| C_{NC}$, showing the decreasing trend with increasing $t_{NC}$. (b) Total normalized drain current **PSD** showing clear reduction with increasing $t_{NC}$ due to the Flicker noise (i.e., the dominant noise source) reduction.

Figure 3(a) shows the matching of Eq. (4) with noise experimental data[22]. The model is able to reproduce the decreasing trend of the Flicker noise with increasing $t_{FE}$, demonstrating the beneficial effect of the ferroelectric layer in terms of noise reduction. The noise contributions due to the electrolyte, ferroelectric and MOSFET channel are found to be negligible compared to the Flicker noise of the transistor, hence they are not a limiting factor for the Signal-to-Noise Ratio ($SNR$). More details regarding the noise sources are found in the supporting information.

The total drain current $PSD$ (normalized to the square of the DC current) is shown in Fig. 3(b), with a clear decreasing trend with increasing $t_{NC}$. The reason for the $PSD$ reduction with increasing $t_{NC}$ is related to the reduction of the Flicker noise, as discussed previously. The increased equivalent oxide capacitance at the denominator in Eq. (4) explains the reduction in the noise, and it is in agreement with experimental observations [see Fig. 3(a)]. The noise reduction provided by the negative capacitor is particularly important for biosensors given the difficulty in providing low noise amplifiers on small geometries due to the high value of the intrinsic flicker noise[35].

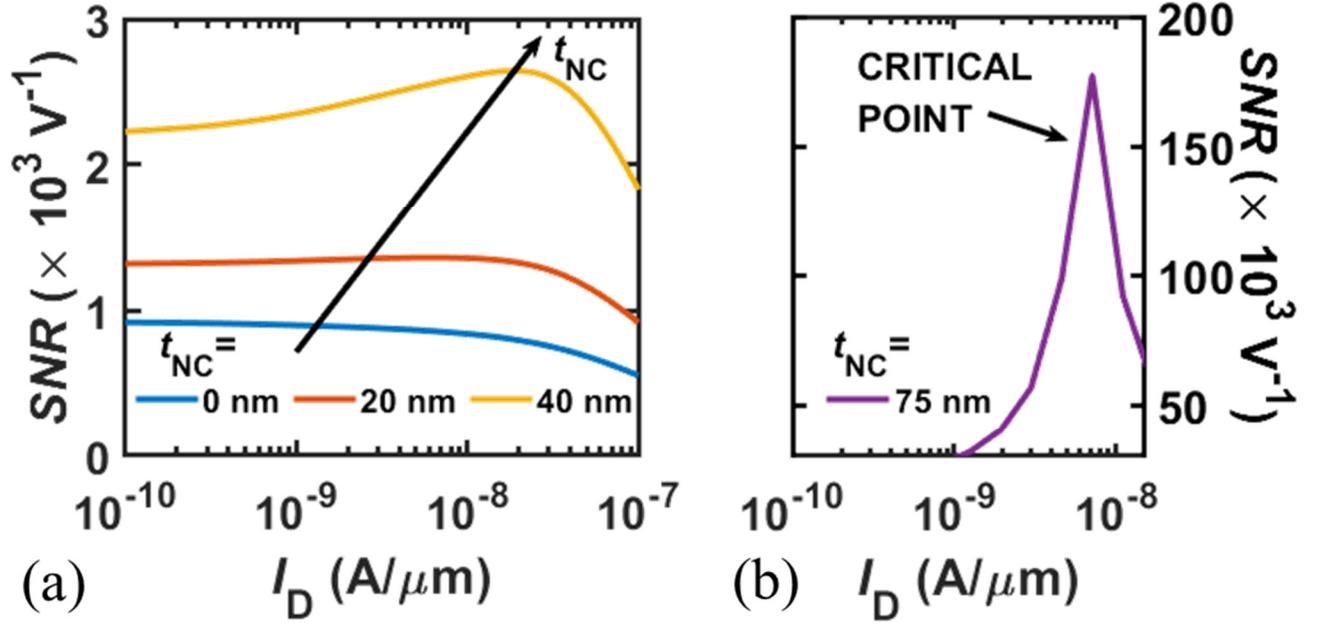

**Figure 4.** Intrinsic *SNR* trend with increasing $t_{NC}$ for (a) small negative capacitor thickness and (b) high $t_{NC}$ (that triggers the abrupt switching of the current at the critical point). The *SNR* shows qualitatively an opposite behavior to the normalized *PSD* (even for small $t_{NC}$) which can be attributed to the Flicker noise reduction. Note that for $t_{NC}$ = 75 nm, the *SNR* peaks at the peak of $g_m$.

The Flicker noise reduction, along with the sensitivity increase, leads to the improved *SNR* defined as following[36]:

$$SNR = \frac{\Delta I_D}{\delta I_{D,n}} \qquad (5)$$

where $\Delta I_D$ is the difference between the current before and after the capture of biomolecules, and $\delta I_{D,n}$ is the noise signal superimposed to the DC value, calculated from the total noise *PSD* as $\delta I_{D,n} = \sqrt{\int PSD_{I_D} df}$. Note that the *SNR* can be more conveniently expressed per unit voltage (considering that $\Delta I_D \sim g_m \Delta V_{FG}$ and normalizing by $\Delta V_{FG}$) to have a quantitative expression for the *intrinsic SNR*, independent of the input signal. Figure 3(d) shows that the increase in $t_{NC}$ improves the *SNR* for all regions of operation. For $t_{NC}$ = 75 nm, the NC-BioFET reaches a maximum *SNR* of ~2×10⁵ V⁻¹ that is almost two orders of magnitude higher than that of a Silicon Nanowire BioFET[36]. We mention the fact that, although in this work the ferroelectric thickness required to have a significant improvement of sensitivity (and thus $SNR$) is relatively high, this should be not a concern in biosensors where normally the gate length is not as small as in digital applications[12,26]. Nonetheless, it is possible to reduce the ferroelectric thickness in NC-BioFETs and still obtain the same benefits in terms of $SNR$ by properly doping the HfO$_2$ layer; for example, as shown in the supporting information, by using Si instead of Zr as the doping atoms for HfO$_2$ it is possible to use a ferroelectric with $t_{NC} = 16$ nm and obtain similar maximum $SNR$ to the case with HZO. This result shows that the NC-BioFETs can be successfully scaled to realize nano-biosensors with enhanced $SNR$.

To summarize, in this letter we have proposed the concept of NC-BioFETs, a new class of biosensors exploiting the negative capacitance effect to improve the Signal-to-Noise Ratio ($SNR$). As the baseline device, we considered the 2D semiconductor FET given its applicability for next-generation ultra-scaled biosensors. We found that, upon the triggering of the non-linear effects associated with the negative capacitance, the sensitivity of NC-BioFETs is improved for a thick enough ferroelectric layer. At the same time, Flicker noise of the transistor is reduced as a result of the increase in the equivalent gate oxide capacitance, thereby improving the $SNR$. Finally, by properly choosing the dopant atoms for the ferroelectric HfO$_2$, it is possible to reduce the ferroelectric thickness while still obtaining the instability required to improve $SNR$, thereby enabling the scalability of NC-BioFETs with current CMOS processes.

## ASSOCIATED CONTENT

**Supporting Information**. The supporting information include the calibration curves for the 2D FET characteristics with experimental data, details regarding the modeling approach and simulation results with the HSiO ferroelectric oxide (PDF).

## AUTHOR INFORMATION

**Corresponding Author**

*E-mail: alam@purdue.edu.

**Present Addresses**

†N. Zagni is now with the Department of Engineering "Enzo Ferrari", University of Modena and Reggio Emilia, Modena (Italy).

**Author Contributions**

M.A.A. conceived the idea. N. Z. developed the simulation framework and carried out the numerical simulations. N.Z., P.P. and M. A. A. co-wrote the manuscript and all authors commented on it.

**Notes**

The authors declare no competing financial interest.

**ACKNOWLEDGEMENTS**

The authors would like to thank Kamal Karda (Purdue University) for useful discussion.

**TABLE OF CONTENTS GRAPHIC**

This figure should be placed beside the abstract as a Table of Contents graphic.

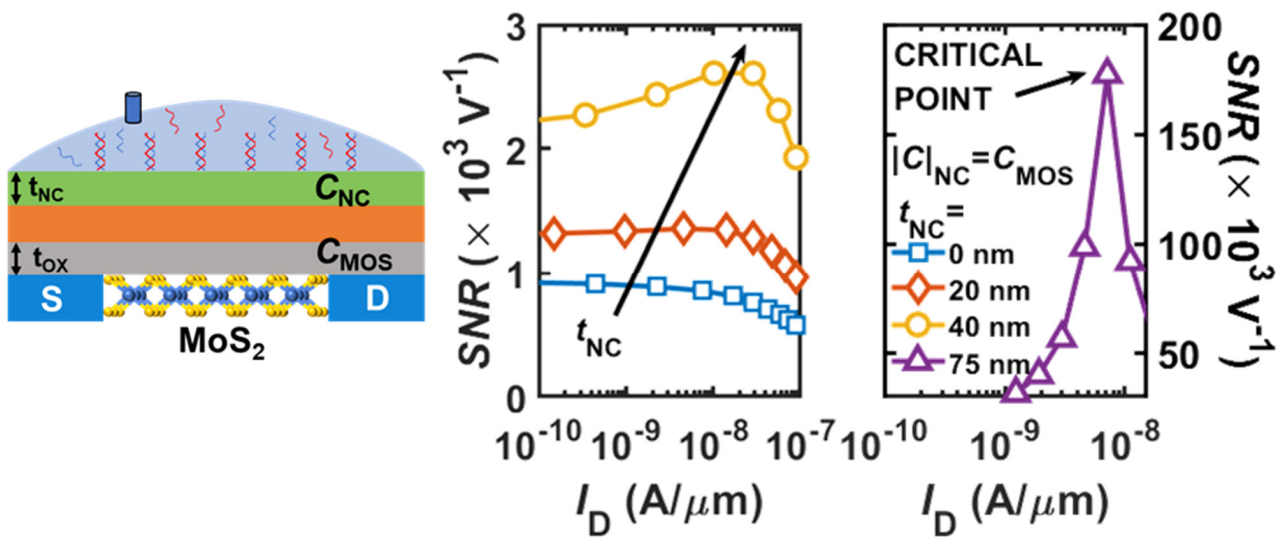

# Supporting Information: Two-Dimensional MoS$_2$ Negative Capacitor Transistors for Enhanced (Super Nernstian) Signal-to-Noise Performance of Next-generation Nano Biosensors

Nicolò Zagni[1,2], Paolo Pavan[2], Muhammad A. Alam[1,*]

[1] *School of Electrical and Computer Engineering, Purdue University, West Lafayette, Indiana 47907, USA*

[2] *Department of Engineering "Enzo Ferrari", University of Modena and Reggio Emilia, Modena, 41125, Italy*

The supporting information includes:

1. The calibration curves of the MoS$_2$ NCFET underlying the NC-BioFET.
2. Details regarding the compact model of the sensor and the NCFET.
3. Details regarding the noise contributions to the overall power spectral density.
4. The results of the simulations with HSiO as the ferroelectric layer.

**Calibration of the Simulations**

The calibration of the MoS$_2$ NCFET compact model with experimental results obtained from the literature[1] is shown in Fig. S1. Calibration includes both $I_D - V_{GS}$ [Fig. S1(a)] and $I_D - V_{DS}$ [Fig. S1(b)] characteristics. The calibration is required for validating the Flicker noise model used to explain the reduction with increasing negative capacitance layer thickness. The parameters used for the model are collected in Table SI. The parameters used to fit the experimental P-V characteristic of the negative capacitance layer (HZO) are collected in Table SII, and are taken from the literature[2].

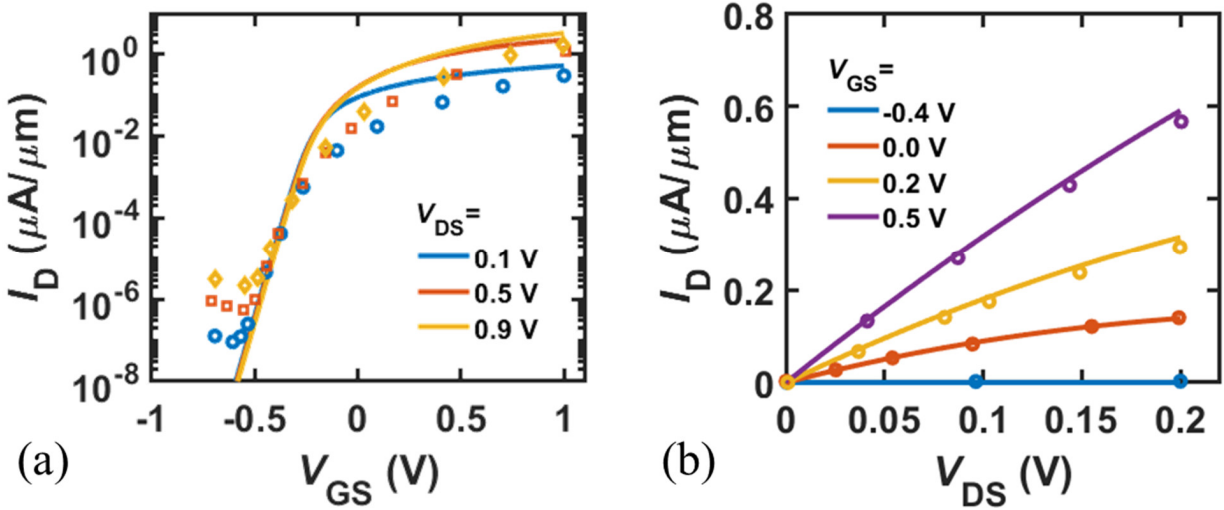

**Figure S1**. Calibration of the MoS$_2$ NCFET compact model with experimental results from the literature[1]. (a) $I_D - V_{GS}$ and (b) $I_D - V_{GS}$ characteristics.

**Table SI**. MoS$_2$ NCFET parameters used for the calibration[1].

| Parameter | Symbol | Value |
| --- | --- | --- |
| Gate Length | $L_G$ | 2 µm |
| Gate Width | $W_G$ | 1 µm |
| Channel Thickness – MoS$_2$ | $t_{CH}$ | 10 nm |
| Oxide Thickness – Al$_2$O$_3$ | $t_{ox}$ | 2 nm (EOT ~ 0.98 nm) |
| Areal Doping Concentration | $N_d$ | $1.5 \times 10^{12}$ cm$^{-2}$ |
| Parasitic Capacitance | $C_P$ | 3.54 fF/µm |
| Mobility | $\mu_n$ | 2 cm$^2$/V.s |
| Ferroelectric Thickness - HZO | $t_{NC}$ | 20 nm |

**Table SII**. LK Parameters for the two ferroelectric layers under study, HZO and HSiO[2,3].

| Parameter | Value (HZO) | Value (HSiO) |
| --- | --- | --- |
| α | $-1.911 \times 10^8$ m/F | $-8.660 \times 10^8$ m/F |
| β | $5.898 \times 10^9$ m$^5$/F/C$^2$ | $1.925 \times 10^{10}$ m$^5$/F/C$^2$ |
| γ | 0 m$^9$/F/C$^4$ | 0 m$^9$/F/C$^4$ |
| ρ | $1.8 \times 10^{-3}$ Ω.m† | |

†Determined by Fourier Infrared spectroscopy analysis for HfO$_2$[4].

**Modeling**

In the following, we describe the fundamental blocks of the compact model more in detail. The BioFET sensor general working principle is analogous to that of the well-known ion-sensitive FET (ISFET)[5], i.e., a MOSFET device that has the metal gate replaced by an external reference electrode immerged in an electrolytic solution. The solution contains the target charged analytes (e.g., DNA, protein, etc.) that are detected when they conjugate with the receptors sites on the insulator surface. The capture of the charged molecules causes a modulation in the surface potential of the MOSFET, which in turn leads to modulations in the drain current. The interaction of charges at the interface between the electrolyte and the oxide of the MOSFET gives rise to the so-called double-layer charge, $Q_{dl}$, which alters the potential between the reference electrode and the gate of the transistor depending on the amount of charge interaction. In addition, a surface charge, $Q_{surface}$, builds-up as a result of the protonation and deprotonation reactions in the electrolytic solution, contributing to screening. These processes can be modeled as an equivalent voltage dependent source for the purpose of circuit simulations[6].

The MoS$_2$ NCFET is modeled by considering the baseline FET and the negative capacitance separately, solved self-consistently by imposing the Kirchhoff's voltage law (KVL):

$$V_{GS} = V_{NC} + V_{MOS} \tag{S6}$$

where $V_{GS}$ is the total gate voltage, $V_{NC}$ is the voltage drop across the negative capacitance layer, and $V_{MOS}$ is the voltage seen by the gate of the baseline MoS$_2$ FET. The analytical expressions for $I_D$ and $Q_{MOS}$ can be found in ref.[2].

**Noise Contributions**

The four different noise contributions considered in this work are shown in Fig. S2. We find that the thermal noise coming from the conductive electrolyte, Fig. S2(a), and the ferroelectric, Fig. S2(b), increase as $t_{NC}$ increases. The increase is explained by the fact that the negative capacitor noise is directly proportional to $t_{NC}$, see equation (2) in the paper, whereas the electrolyte thermal noise increases due to the amplification given by the negative capacitor. As expected, the MOSFET channel noise, Fig. S2(c), does not vary with $t_{NC}$, as it is due to the conduction mechanism in the semiconductor that is not affected by the presence of the ferroelectric. All these noise contributions are negligible compared to the Flicker noise of the transistor, Fig. S2(d), hence they are not a limiting factor for the Signal-to-Noise Ratio ($SNR$). Flicker noise is indeed found to be order of magnitudes higher than the other noise sources in this work. We ascribe this to the relatively high defect density (~$10^{21}$ eV$^{-1}$cm$^{-3}$) necessary to reproduce the experimental data as shown in Fig. 3(a) of the paper.

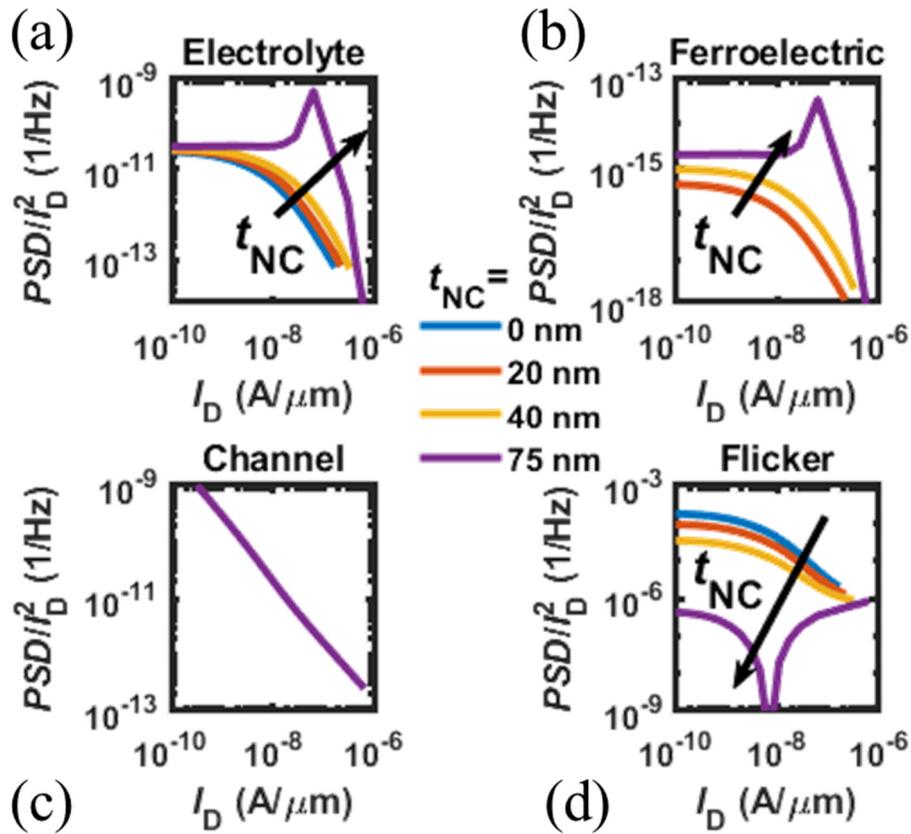

**Figure S2**. Noise contributions to the total power spectral density: (a) electrolyte, (b) ferroelectric, MOSFET (c) channel and (d) Flicker noise. While electrolyte and ferroelectric noise increase, channel noise remains unvaried and Flicker noise reduces with increasing $t_{NC}$.

**HSiO versus HZO results**

To demonstrate the potential for integration of the NC-BioFET, we analyze the device with another ferroelectric layer, namely Si- doped HfO$_2$ (HSiO)[7]. With HSiO, the instability caused by the negative capacitance can be obtained with lower $t_{NC}$ compared to HZO. This stems from the weaker capacitance obtained with HSiO at the same $t_{NC}$ as explained as following. In a first-order approximation, the first term of equation (2) in the paper allows interpreting $|\alpha|$ as the inverse of the dielectric constant of a conventional capacitance ($C \sim \varepsilon/t$). Hence, a larger $\alpha$ (as in the case of HSiO compared to HZO) leads to a lower negative capacitance (in absolute value).

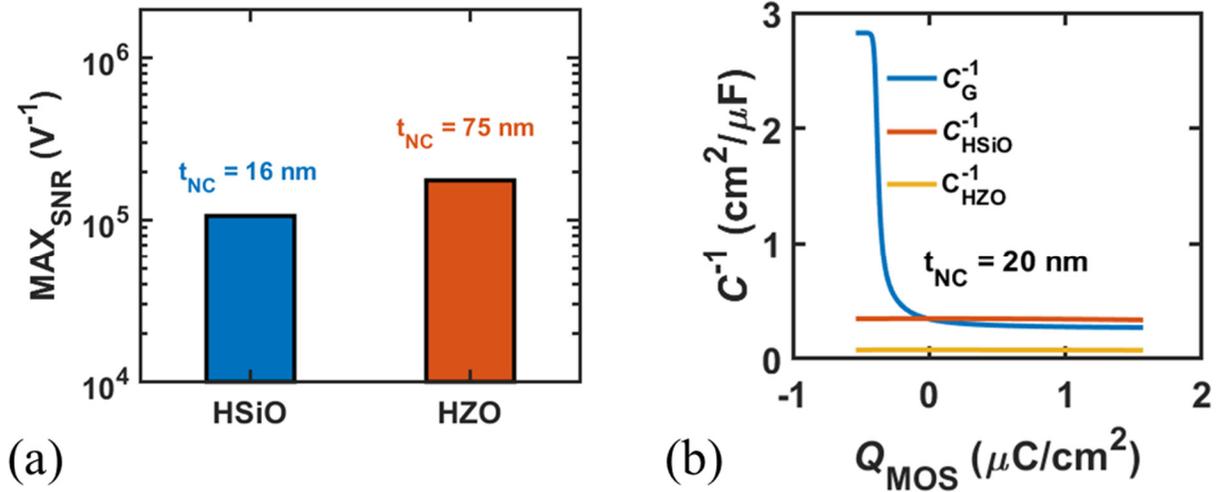

**Figure S3**. (a) Comparison of the simulation results obtained by replacing HZO with HfSiO, showing that the same benefits of the NC-BioFET can be obtained with lower negative capacitance layer thickness, thereby allowing a more aggressive scaling. The table in the inset shows the parameters of the HfSiO ferroelectric[7]. (b) Comparison of the negative capacitance obtained with HSiO and HZO with the MOSFET capacitance $C_G$ showing that for example, with **20 nm** thickness the matching condition is achieved only with the HSiO and not with HZO.

The comparison between the results obtained with the two different ferroelectric materials is shown in Fig. S3(a), which confirms the fact that an appreciable non-linearity already arises at 16 nm with HSiO, compared to 75 nm with HZO (see Table SII for the ferroelectric parameter values). Figure S3(b) shows that at a given $t_{NC}$, $C_{HSiO}$ is lower than $C_{HZO}$ hence guaranteeing the matching condition at lower thickness. We conclude that by properly choosing the dopant species for HfO$_2$ it is possible to realize ultra-scaled negative capacitance transistors that can be employed for the enhancement of SNR in BioFETs.